\documentclass[twocolumn,preprintnumbers,,aip,apl]{revtex4}

\usepackage{graphicx,amsmath,amssymb}

\begin{document}

\title{Real-space-transfer mechanism of negative differential conductivity in gated graphene-phosphorene hybrid structures: Phenomenological heating model}

\author{V. Ryzhii$^{1,2,3}$,
 M. Ryzhii$^4$,  D. Svintsov$^5$,  V. Leiman$^5$, 
P. P. Maltsev$^2$,
D. S. Ponomarev$^2$,\\
 V. Mitin$^{6}$,  M. S. Shur$^{7,8}$, and T. Otsuji$^1$, }
\affiliation{
$^1$ Research Institute of Electrical Communication, Tohoku University,  Sendai 980-8577, Japan\\
$^2$ Institute of Ultra High Frequency Semiconductor Electronics of RAS, Moscow 117105, Russia\\
$^3$ Center for Photonics and Infrared Engineering, Bauman, Moscow State Technical University, Moscow 111005, Russia\\
$^4$ Department of Computer Science and Engineering, University of Aizu, Aizu-Wakamatsu, 965-8580, Japan\\
$^5$ Laboratory of 2D Material's Optoelectronics, Moscow Institute of Physics and Technology, Dolgoprudny 141700, Russia\\
$^6$ Department of Electrical Engineering, University at Buffalo, SUNY, Buffalo, NY 1460-1920, USA\\
$^7$ Departments of Electrical, Electronics, and Systems Engineering and Physics, Applied Physics, and Astronomy, Rensselaer Polytechnic Institute, Troy, NY 12180, USA\\
$^8$ Electronics of the Future, Inc., Vienna, VA 22181, USA\\ 
}

%

\begin{abstract}
We analyze the nonlinear carrier transport in the gated graphene-phosphorene (G-P) hybrid structures - the G-P field-effect transistors (G-P-FETs)  using a phenomenological   model. This model assumes that due to high carrier densities  in the G-P-channel, the carrier system, including the electrons and holes in both
the G- and P-layers, is characterized by a single effective temperature. We demonstrate that a strong electric-field dependence of the G-P-channel conductivity and  substantially non-linear current-voltage characteristics,
exhibiting a negative differential conductivity, are associated with the carrier heating and the real-space carrier  transfer
between the G- and P-layers. The predicted features of the G-P-systems can be used in the  detectors and sources of electromagnetic  radiation   and in the logical circuits. 
\end{abstract}

\maketitle
\newpage 
\section{Introduction}

Unique properties of Graphene (G)~\cite{1} and  recent advances in technology of van der Waals materials~\cite{2,3} present an excellent opportunity for developing effective electronic and optoelectronic devices~\cite{3,4,5,6,7,8,9,10,11,12,13,14,15}. Combining the G-layers with the gapless energy spectrum and enhanced electron (and hole) mobility
  and a few-layer black phosphorus layer or phosphorene (P)~\cite{16,17,18,19,20,21,22,23} exhibiting the flexibility of the band structure, open up remarkable prospects for the creation of novel devices, in particular, photodetectors~\cite{24}.

The G-P hybrid structures can be used in the real-space-transfer (RST) devices. 
The RST devices exhibiting the negative differential conductivity (NDC) have attracted a lot of attention since their proposal by Z. S. Gribnikov in early 1970's~\cite{25} and further developments~(for example,~\cite{26,27,28,29,30,31}, see also Refs.~\cite{32,33} and references therein). The RST devices  exhibit interesting features including high speed operation.
Most of the  RST devices have been based on the A$_3$B$_5$ heterostructures. Their operation is associated with the electric-field heating of the electron gas in the narrow-gap channel (in particular, GaAs channel), resulting in the transfer of the hot relatively light electrons to an adjacent wide-gap layer with a higher electron effective mass (such as an  Al$_x$Ga$_{1-x}$As-layer). A decrease in   the fraction of the light electrons accompanied with an increase of the heavy electrons with an increasing electric field results in the roll-off of the net electron system conductivity and,  could lead to  the NDC. 
In this paper, we propose
to use the effect of the RST in the gated graphene-phosphorene (G-P) hybrid sandwich-like structures, i.e., in the  
G-P-channel field-effect transistors (G-P-FETs) 
 and evaluate the characteristics
of such devices. In contrast to the effect of NDC in the standard semiconductors (in which the net electron or hole density does not markedly vary even at a strong heating), in the G-P- channels  the net carrier density can
be pronouncedly changed. This adds a substantial complexity to the operation the G-P-channel devices.   

\section{Model}

  Figure~1 demonstrates a schematic view of the gated G-P structure (i.e. a FET  with the G-P channel). It is assumed that the P-layer  consisting of a few atomic layers is oriented in such a way that the direction from the FET source to its drain  corresponds to the zigzag direction.
The dynamics of electrons and holes in this direction is characterized by a large effective mass.
As a result, the RST of the electrons  and holes from the G-layer (where their mobility can be very high) to the  P-layer (with
relatively low mobility in the direction in question) can enable  sharp current-voltage characteristics with an elevated peak-to-valley ratio (i.e. a large absolute value of the NDC). 
The selection of the P-layer for the hybrid structure under consideration is associated not only with a high electron and hole  masses (and low mobility) in the electric-field direction, but also with a wide opportunity to provide
desirable height of the barrier between the P-layer conduction band bottom and the Dirac point in the G-layer.
This can be realized by a proper choice of the number of  atomic layers in the P-layer~\cite{16,17}.

 For the sake of definiteness, we consider the P-layer consisting of a few atomic P-layers ($N= 2 -3$), assuming that it is generally doped (the pristine P-layers are of p-type).
 The gate voltage $V_G$ can substantially vary the carrier densities  in the G-P-channel, so that
 the latter comprises the two-dimensional electron and hole gases in  both the G- and P-layers.

The band gap $\Delta$ and the energy spacing $\Delta^e$ and $\Delta^h$,
between the Dirac point in the G-layer and edges 
of the conduction and valence bands,  (determined by the pertinent work functions) depend on the number $N$.
In the heterostructure under consideration,  $\Delta^e \sim \Delta^h \simeq 0.4 -  0.45$~eV ($N = 3$) ~\cite{16}. 

To calculate  the G-P FET current-voltage characteristics and evaluate the NDC, as a first step, we use a simplified semi-classical phenomenological model for the electrons and holes in such a FET channel, although for more rigorous treatment
of the RST,
the quantum approach is needed~\cite{34,35,36}.

\begin{figure}[t]
\centering
\includegraphics[width=7.0cm]{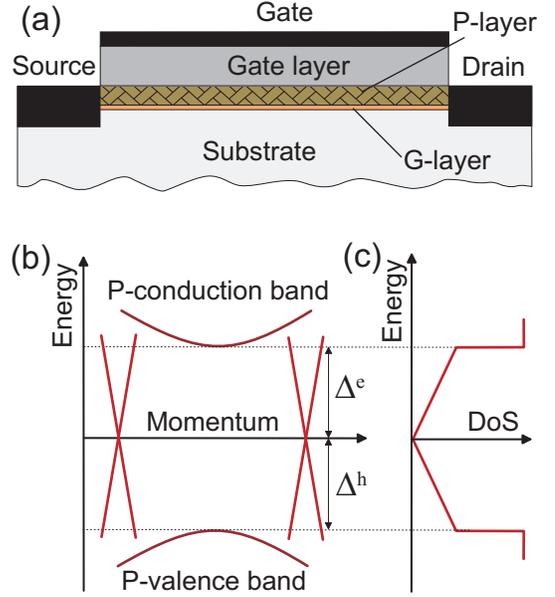}
\caption{Schematic view of (a) the G-P FET structure, (b) energy spectrum, and (c) density of states (DoS). Cone-shaped parts of the energy spectrum correspond to the K- and K$^{\prime}$ - valleys in G-layers.}
\label{F1}
\end{figure}

We set  the dispersion relations for the electrons and holes in the G- and P-layers as follows:

\begin{equation}\label{eq1}
\varepsilon_G^e = v_W\sqrt{p_x^2 + p_y^2}, \qquad \varepsilon_P^e = \Delta^e + \frac{p_x^2}{2m_{xx}} + 
\frac{p_y^2}{2m_{yy}} ,
\end{equation}

\begin{equation}\label{eq2}
 \varepsilon_G^h = -v_W\sqrt{p_x^2 + p_y^2}, \qquad  \varepsilon_P^h = - \Delta^h - \frac{p_x^2}{2m_{xx}} - \frac{p_y^2}{2m_{yy}},
\end{equation}
respectively. Here $v_W \simeq 10^8$~cm/s  is the characteristic velocity of electrons in the G-layers,
$m_{xx} = M$ and $m_{yy} =m$ are the components of the effective mass tensor ($M \gg m$), $p_x$ and $p_y$
are the carrier momenta in the source-drain direction and the perpendicular direction, respectively.
The components of the effective mass tensor for both conduction and valence bands for $N = 2 - 3$ are approximately as follows:
$m  \simeq 0. 06m_0$ and $M \simeq 1.35m_0$ ($N = 3$), where $m_0$ is the mass of bare electron.

Figure~1 shows schematically the G-P-FET structure, the energy spectrum of electrons and holes, and the pertinent energy dependence of the density of states, corresponding to Eqs.~(1) and (2).

At a relatively high  frequency of the electron-electron, hole-hole, and electron-hole  collisions, the electron-hole plasma  in the G-P channel
can be characterized by its common electron effective temperature $T$ (in the energy units, equal for both
the G- and P-layers) but generally different quasi-Fermi energies $|\mu^e| \neq |\mu^h|$, so that the electron and hole distribution functions are $f^{e}(\varepsilon) = \{\exp[(\varepsilon - \mu^{e})/T] + 1\}^{-1}$ and 
$f^{h}(\varepsilon) = \{\exp[(\varepsilon - \mu^{h})/T] + 1\}^{-1}$, respectively.

The model under consideration is based on the following assumptions:

(1) The sufficiently frequent inter-carrier  collisions enable the establishment of the quasi-Fermi distributions with 
the common effective temperature, $T$ for all electron and hole components in both G- and P-layers. Sufficiently strong  interactions between the electrons (and holes) belonging to neighboring layers promotes the inter-layer equilibrium~\cite{37,38,39}.  

(2) Due to heavy electron and hole effective masses $M$ and $\sqrt{mM}$, 
the conductivity of the P-layer is relatively small because this layer mobility in the direction corresponding to the mass $M$ is proportional to $ 1/\sqrt{mM}M$~\cite{40}.
Hence 
we disregard the P-layer conductivity in comparison with
the G-layer conductivity. The same assumption is valid if the P-layer is disconnected from the source and drain contacts. We also neglect the contribution of the heavy carriers to the energy balance.\\

(3) The  momentum relaxation of the electrons and holes in the G-layer   (light electrons and holes) is due to their scattering on defects,  impurities, acoustic phonons,  and the heavy electrons and holes in the P-layer.
The latter and the charged impurities are assumed to be screened (see Sec.~V).
We believe that
the energy relaxation at the room temperature under consideration is associated with optical phonons in the G-layer,
and the interband transitions assisted by the optical phonons are  the  main recombination-generation mechanisms, neglecting the Auger generation-recombination processes.  Due to the prohibition of the Auger processes in  the G-layers with the ideal linear gapless energy spectrum because of the energy and momentum conservation laws~\cite{41}, 
even 
in non-ideal G-layers (see~\cite{42} and references therein),  there is an ambiguity of  the characteristic times ratio, $\tau_0^{inter}/\tau^{inter}_{Auger}$,
of the interband transitions mediated by the optical phonons and electron-hole (Auger) processes,
particularly, at different carrier temperatures. The characteristic time  $\tau^{inter}_{Auger}$ can be fairly different depending on the dielectric constant of the substrate $\kappa$ and the spacing between the G-P-channel and the gate $W_g$~\cite{42}.  The case of $\tau_0^{inter} \ll \tau^{inter}_{Auger}$, which is under consideration in the following,  can  conditionally correspond to  a large $\kappa$ and a small $W_g$.

(4) At relatively short  the characteristic times of the electron-hole generation-recombination associated with the optical phonons ,
  the quasi-Fermi energies (counted from the Dirac point) of the electron and hole components can be 
generally  different ($\mu^e \neq -\mu^h$).

\section{Main equations of the model}

\subsection{Conductivity of the G-P channel}

Considering the above model, we use the following formula
 for the net conductivity of the G-P channel $\sigma$  (see, in particular,~\cite{43,44,45,46,47,48,49,50,51,52,53}):

\begin{eqnarray}\label{eq3}
 \sigma = -\frac{e^2T_0\tau_0}{\pi\hbar^2}\biggl(\frac{T}{T_0}\biggr)^{l+1}\frac{\Sigma_I}{(\Sigma_I + \Sigma_{P})}\nonumber\\
\times \int_0^{\infty}d\xi\xi^{l+1}\frac{d}{d\xi}[f^e (\xi) + f^h(\xi)].
\end{eqnarray}
Here $\tau(p) = \tau_0(pv_W/T_0)^l = \tau_0(T/T_0)^l\xi^l$ is the momentum relaxation time,  $\xi = pv_W/T_0$ is the normalized сarrier energy in the G-layer, $\Sigma_{I}$ is the density of the  scatterers, $\Sigma_P(T)$ is the carrier density in the P-layer at the effective temperature $T$ (which, due to their large effective mass, can also be considered as the effectively screened Coulomb  scatterers for the light electron and holes in the G-layer), and  $\hbar $ is the Planck constant. If the weakly screened Coulomb scattering prevails,  $l = 1$ (for example,~\cite{43,45,46,50,52}).
In the case of dominant scattering on neutral scatterers, $l = -1$.  

Due to the above assumptions, in the following we set $l = -1$.
 In this case, Eq.~(3) yields

\begin{equation}\label{Eq4}
 \sigma  = \sigma_0 [f^e(0) + f^h(0)] \frac{\Sigma_I}{(\Sigma_I + \Sigma_{P})}.
\end{equation}
Here $\sigma_0 =  (e^2T_0\tau_0/\pi\hbar^2)$ is the characteristic conductivity (it is equal to the low electric-field conductivity in the case of neutral scatterers).

\subsection{Carrier interband balance}

The carrier densities, $\Sigma_G$ and  $\Sigma_P$, in the G- and P-layer 
are, respectively,  given by
\begin{eqnarray}\label{eq5}
 \Sigma_{G} 
= \Sigma_0\biggl(\frac{T}{T_0}\biggr)^2\biggl[{\cal F}_{1}\biggl(\frac{\mu^e}{T}\biggr)  + {\cal F}_{1}\biggl(\frac{\mu^h}{T}\biggr)\biggr],
\end{eqnarray}

\begin{eqnarray}\label{eq6}
 \Sigma_{P} 
= \Sigma_N\ln\biggl\{\biggl[1 +\displaystyle\exp\biggl(\frac{\mu^e -\Delta^e}{T}\biggr)\biggr]
\biggl[1 +\displaystyle\exp\biggl(\frac{\mu^h -\Delta^h}{T}\biggr)\biggr]\biggr\},
\end{eqnarray}
 where 
$$
{\cal F}_1(a)
 =\int_0^{\infty}\frac{d\xi\xi}{\exp(\xi - a) + 1}
$$
is the Fermi-Dirac integral~\cite{54}, $\Sigma_0 = 2T_0^2/\pi\hbar^2\hbar^2$, and  $\Sigma_N = N T_0\sqrt{mM}/\pi\hbar^2$. The factor $N$ in the latter formula reflects the fact that  the density of states in the few-layer P-layer scales roughly with the layer number $N$~\cite{38}.
 
The net surface charge  density in the G-P- channel   $e\Sigma = \kappa\,|V_G - V_{CNP}|/4\pi\,W_g$ induced by the applied gate voltage $V_G$  
comprises the electron,  $\Sigma_G^e$ and $\Sigma_P^e$, and hole, $\Sigma_G^h$ and $\Sigma_P^h$ densities.
Here $V_g = V_G - V_{CNP}$ 
the gate-voltage swing,  $V_{CNP} \propto \Sigma_I^{ch}$  is the voltage, which 
corresponds to the charge-neutrality point, $\Sigma_{I}^{ch}$
is the density of non-compensated charged impurities, $\kappa$ and $W_g$ are the background dielectric constant and the thickness of the gate layer, respectively, and $e$ is the electron charge. 
%
%
Considering the above, the gate voltage swing $V_g$ and  the quantities $T$, $\mu^e$ and $\mu^h$ are related to each other as
\begin{eqnarray}\label{eq7}
 \frac{V_g}{V_0} =\biggl(\frac{T}{T_0}\biggr)^2
 \biggl[{\cal F}_{1}\biggl(\frac{\mu^e}{T}\biggr)  - {\cal F}_{1}\biggl(\frac{\mu^h}{T}\biggr)\biggr]
 \nonumber\\
+
\gamma_0 \frac{T}{T_0}\ln\frac{\biggl[1 +\displaystyle\exp\biggl(\frac{\mu^e -\Delta^e}{T}\biggr)\biggr]}{
\biggl[1 +\displaystyle\exp\biggl(\frac{\mu^h -\Delta^h}{T}\biggr)\biggr]},
\end{eqnarray}
Here 
$\gamma_0 = N\sqrt{mM}v_W^2/2T_0$,  $V_0 = 8eT_0^2W_g/\kappa\hbar^2v_W^2$,  and $N$ is the number of the monolayers in the P-layer (for a moderate $N$).
The parameter $\gamma_0$ can be large  due to a relatively high density of states
in the P-layer.

In the limit $T = 0$ at not too high gate voltage swing $V_G - V_{Dirac}$ when the P-layer is empty, Eq.~(7) yields the standard expression for the Fermi energy degenerate electron gas in the G-layer: $\mu = \hbar\,v_W\sqrt{\pi\Sigma}$.
If $\Sigma_N = 0$ at very high electron and hole effective temperatures, $\mu$ tends to zero as $\mu \propto T^{-1}$.

In the following for definiteness and simplicity we set $\Delta^e = \Delta^h = \Delta/2$ (this is approximately valid for $N = 2$ and $N =3$~\cite{16}).

\begin{figure}[b]
\centering
\includegraphics[width=7.0cm]{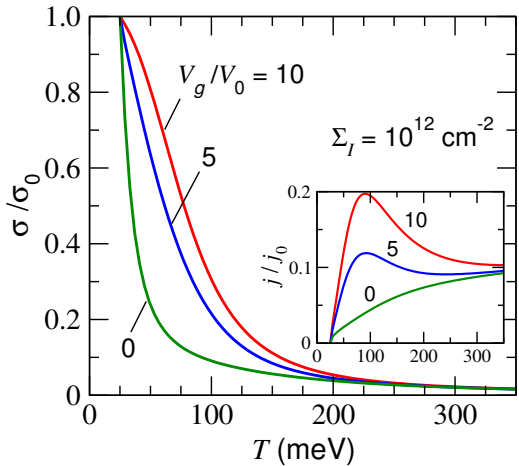}
\caption{Normalized conductivity vs effective temperature. The inset shows the pertinent temperature dependence of the normalized current density.}
\label{F2}
\end{figure}

\begin{figure}[b]
\centering
\includegraphics[width=7.0cm]{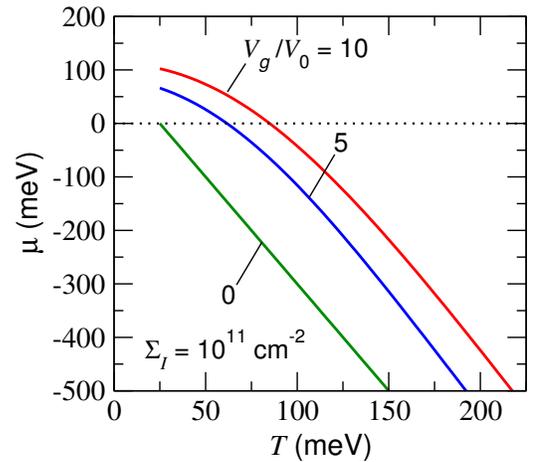}
\caption{Quasi-Fermi energy $\mu$  vs effective temperature for $\Sigma_{I} = 10^{11}$~cm$^{-2}$ at different normalized gate-voltage swings $V_g/V_0$.}
\label{F3}
\end{figure}

\subsection{Generation-recombination and energy balance equations}

The equation governing the interband balance of the carriers can be generally presented as
\begin{equation}\label{eq8} 
G_{Auger} + G_{op} + G_{ac} + G_{rad} = 0,
\end{equation}
where the terms in Eq.~(8) correspond to different processes: the interband Auger generation-recombination  processes
 and the processes associated with optical-phonon, acoustic-phonon, and radiative transitions (in particular, indirect transitions
 for which some selection restrictions are lifted). 
 At sufficiently  fast processes of the optical phonon decay into acoustic phonons followed by their effective
removal,  that  is confirmed by high values of the G-layer thermal conductivity~\cite{55,56,57,58},
 the optical phonon system in the heterostructures under consideration is in equilibrium
with the thermal bath  with the temperature $T_0$. 
 
For the optical-phonon term(under a  rather natural assumption $\hbar\omega_0 > |\mu^e|, |\mu^h|, T$), we use the following expression~\cite{59,60} (see also ~\cite{61}):

\begin{equation}\label{eq9}
G_{op} = \frac{\Sigma_0}{\tau_0^{inter}}
\biggl[{\cal N}_0 - ({\cal N}_0 + 1)\exp\biggl(\frac{\mu^e + \mu^h - \hbar\omega_0}{T}\biggr)\biggr].
\end{equation}
Here ${\cal N}_0 = [\exp(\hbar\omega_0/T_0) - 1]^{-1} \simeq \exp(-\hbar\omega_0/T_0)$ is the number of equilibrium optical phonons in the G-layer with the energy $\hbar\omega_0$, 
$ (\Sigma_{0}/\tau_0^{inter})\exp(-\hbar\omega_0/T_0) = G_0^{eq}$, 
where $G_0^{eq}$ is the rate of the thermal generation of the electron-hole pairs in the G-layers due to absorption of optical phonons in equilibrium (the latter was estimated as $G_0^{eq} \simeq 10^{21}$~cm$^{-2}$s$^{-1}$ at $T_0 = 300$~K~\cite{59,60,61}), 
$\tau_0^{inter}$ is the characteristic time of the spontaneous optical phonon emission accompanied at the  interband transitions.

In the situation under consideration, we present the energy balance equation in the following form:

\begin{eqnarray}\label{eq10}
\frac{\hbar\omega_0\Sigma_{0}}{\tau_0^{inter}}\biggl[({\cal N}_0 + 1)\exp\biggl(\frac{\mu^e + \mu^h -\hbar\omega_0}{T}\biggr) - {\cal N}_0\biggr] \nonumber\\
+ \frac{\hbar\omega_0\Sigma_G}{\tau_0^{intra}}\biggl[({\cal N}_0 + 1)\exp\biggl( - \frac{\hbar\omega_0}{T}\biggr) -   {\cal N}_0\biggr] = 
\sigma E^2.
\end{eqnarray}
 Here  
$\tau_0^{intra}$ is the characteristic time of the spontaneous optical phonon emission accompanied with the intraband  electron and hole transitions, $\Sigma_G \propto [{\cal F}_1(\mu^e/T) + {\cal F}_1(\mu^h/T)](T/T_0)^2$ is approximately equal to the net carrier density in the G-layer (a small deviation from the real carrier density is associated with the dependence of  the optical phonon emission and absorption probability on the carrier energy, which, in turn, is due to  the density of state linearity)
 and $E \simeq V_{sd}/L$ is the longitudinal source-to-drain electric field
in the channel ($V_{sd}$ is the  voltage applied between the source and drain contacts and $L$ is the length of the channel).

Due to  relatively high values of $\hbar\omega_0$ compared to $T_0$, 
we  set in the following ${\cal N}_0 \simeq \exp(-\hbar\omega_0/T_0) \ll 1$. 
 Setting $\hbar\omega_0 \simeq 0.2$~eV and $T_0 = 0.025$~eV,  the number of optical phonons is estimated as  ${\cal N}_0
\simeq 3.35\times 10^{-4}$.

As assumed above,  the electron-hole generation-recombination processes are  associated primarily with the optical phonon spontaneous emission,
(the pertinent characteristic time $\tau_0^{inter}$ is much shorter than $\tau_{Auger}^{inter}$ associated with the Auger processes),  so that
the equation governing the electron and hole  balance  acquires the following form:

\begin{equation}\label{eq11}
\exp\biggl(\frac{\mu^e + \mu^h- \hbar\omega_0}{T}\biggr) -  \exp\biggl(-\frac{\hbar\omega_0}{T_0}\biggr)
 = 0.
\end{equation}
Equation~(11), yields

\begin{equation}\label{eq12}
\mu^e + \mu^h =\hbar\omega_0\biggl(1 - \frac{T}{T_0}\biggr).
\end{equation}

\subsection{General set of the  equations}

Considering   Eqs.~(4)- (6) and Eq.~(11).
 we arrive at  the following set of the equations governing the carrier effective temperature, quasi-Fermi energy, conductivity, and the G-P channel current-voltage characteristics:

\begin{widetext}

\begin{eqnarray}\label{eq12}
\frac{V_g}{V_0} 
=  \biggl[{\cal F}_1\biggl(\frac{\mu}{T}\biggr) - {\cal F}_1\biggl(-\frac{\mu}{T} - \hbar\omega_0\biggl(\frac{1}{T_0}-  \frac{1}{T}\biggr)\biggr)
\biggr]
\,\frac{T^2}{T_0^2}\nonumber\\
+
\gamma_0 \biggl(\frac{T}{T_0}\biggr)\biggl\{\exp\biggl(\frac{\mu -\Delta/2}{T}\biggr)-
\exp\biggl[-\frac{\mu  +\Delta/2}{T} - \hbar\omega_0\biggl(\frac{1}{T_0} - \frac{1}{T}\biggr)\biggr]\biggr\},
\end{eqnarray}

\begin{eqnarray}\label{eq13}
\frac{\sigma}{\sigma_0} = \frac{1}{1 + \displaystyle\frac{\Sigma_N}{\Sigma_I}\frac{T}{T_0}\displaystyle\exp\biggl(-\frac{\Delta}{2T}\biggr)\biggl\{\exp\biggl(\frac{\mu}{T}\biggr) + \exp\biggl[-\frac{\mu}{T} - \hbar\omega_0\biggl(\frac{1}{T_0}- \frac{1}{T}\biggr)\biggr]\biggr\}
}
\nonumber\\
\times
\biggl\{ \frac{1}{\displaystyle\exp\biggl(-\frac{\mu}{T}\biggr) + 1} + \frac{1}
{\displaystyle\exp\biggl[\frac{\mu}{T}+ \hbar\omega_0\biggl(\frac{1}{T_0} - \frac{1}{T}\biggr)
\biggr]+ 1} \biggr\}, 
\end{eqnarray}
\begin{equation}\label{Eq14}
\biggl[{\cal F}_1\biggl(\frac{\mu}{T}\biggr) + {\cal F}_1\biggl(-\frac{\mu}{T} - \hbar\omega_0\biggl(\frac{1}{T_0}-  \frac{1}{T}\biggr)\biggr)
\biggr]\cdot\biggl[\exp\biggl(-\frac{\hbar\omega_0}{T}\biggr) - \exp\biggl(-\frac{\hbar\omega_0}{T_0}\biggr)\biggr]\,\frac{T^2}{T_0^2} 
 = 
\frac{\sigma}{\sigma_0}\frac{ E^2}{E_0^2},
\end{equation}

\begin{equation}\label{Eq16}
j = \sigma(E) E.
\end{equation}

\end{widetext}
Here  
\begin{equation}\label{Eq17}
E_0 = \sqrt{\frac{2\hbar\omega_0}{T_0}}\biggl(\frac{T_0}{ev_W\sqrt{\tau_0\tau_0^{intra}}}\biggr)
\end{equation}

The channel current can be normalized by 

\begin{equation}\label{eq18}
j_0 = \sigma_0E_0 = \frac{eT_0^2}{\pi\,v_W\hbar^2}\sqrt{\frac{2\hbar\omega_0}{T_0}\frac{\tau_0}{\tau_0^{intra}}}
\end{equation}

In the particular case when $V_g$ corresponds to the Dirac point ($V_g = V_{CNP}$), the above system of equations can be simplified. Indeed, in such a case, Eq.~(13) results in 

\begin{equation}\label{Eq19}
\frac{\mu}{T} = -\frac{\hbar\omega_0}{2}\biggl(\frac{1}{T_0} - \frac{1}{T} \biggr)
\end{equation}
and, hence, Eq.~(14) becomes as follows:
\begin{eqnarray}\label{eq20}
\frac{\sigma}{\sigma_0} = \frac{1}{1 + \displaystyle\frac{2\Sigma_N}{\Sigma_I}\frac{T}{T_0}
\displaystyle\exp\biggl(-\frac{\Delta}{2T}\biggr)\exp\biggl[ -\frac{\hbar\omega_0}{2}\biggl(\frac{1}{T_0}- \frac{1}{T}\biggr)\biggr]}\nonumber\\
\times\frac{2}{1 + \displaystyle\exp\biggl[ \frac{\hbar\omega_0}{2}\biggl(\frac{1}{T_0}- \frac{1}{T}\biggr)\biggr]}.
\end{eqnarray}
Equation~(20) explicitly demonstrates an increase in the conductivity $\Sigma$ with increasing effective temperature $T$. 

\newpage

\section{Numerical  results}

\begin{figure}[t]
\centering
\includegraphics[width=7.0cm]{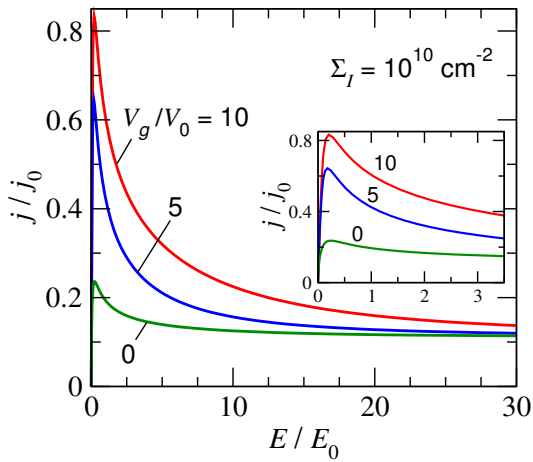}
\caption{Current-voltage characteristics  for $\Sigma_{I} = 10^{10}$~cm$^{-2}$ at different normalized gate-voltage swings $V_g/V_0$.}
\label{F4}
\end{figure}

The set of Eqs.~(13) - (16)   was solved numerically, to obtain the 
effective-temperature and electric-field dependences. The pertinent calculation results are shown in Figs.~2 - 7.
We set  $N = 3$, $\Delta = 900$~meV, $\sqrt{mM} = 0.29m_0 = 2.65\times 10^{-28}$~g,  $\hbar\omega_0 = 200$ meV,
and $T_0 = 25$~meV,
so that $\gamma_0 = 32$, $\Sigma_N = 1.05\times 10^{13}$~cm$^{-2}$. The scatterer density was assumed to be
in the range $\Sigma_I = 10^{10} - 10^{12}$~cm$^{-2}$. The relative gate voltage swing $V_g/V_0$ varied
from zero to ten. The dependences on all  plots  below are normalized by $j_0$ and $E_0$ corresponding to
$\Sigma_I = 10^{11}$~cm$^{-2}$ assuming that $\tau_0 \propto \Sigma_I^{-1}$.

An increase in the electric field leads to a rise of the effective temperature.
Figure~2 shows examples (for $\Sigma_I = 10^{12}$~cm$^{-2}$) of  the  normalized G-P channel conductivity $\sigma/\sigma_0$ as a function of the effective temperature $T \geq T_0 = 25$~meV.  As seen from the plots in Fig.~2, $\sigma/\sigma_0$ exhibits a steep drop with increasing $T$  at all $V_g/V_0$.  At smaller values of $\Sigma_I$,  $\sigma/\sigma_0$ versus $T$ relation  becomes even  steeper.
However,  as shown in the inset in Fig.~2, 
the $j/j_0 - T$ relation, found as an example for $\Sigma_I = 10^{11}$~cm$^{-2}$, can be qualitatively different 
depending on $V_g/V_0$. Figure~3 shows that  the quasi-Fermi $\mu$ also exhibits a steep drop when $T$ increases
(for $V_g  >0$). Moreover, $\mu$ changes its sign at certain values of $T$ depending on  $V_g/V_0$.
This implies that  the electron gas in the G-layer being degenerate at a moderate heating, i.e,
at weak electric fields, becomes nondegenerate with increasing effective temperature.
 The variations of the effective temperature and the quasi-Fermi energy  markedly affect the distribution of the carriers between the G- and P-layers and, hence, the current-voltage characteristics.

Figures~4 - 6 demonstrate the  current-voltage characteristics of the G-P channels with different scatterer density
at different gate voltages.
One can see that the shape of the current-voltage characteristics, in particular, the height of the current peaks and
the peak-to-valley ratio,  are different in the samples with different scatterer densities $\Sigma_I$. Although, the NDC preserves  
when $\Sigma_I$ varies in rather wide range.

\begin{figure}[t]
\centering
\includegraphics[width=7.0cm]{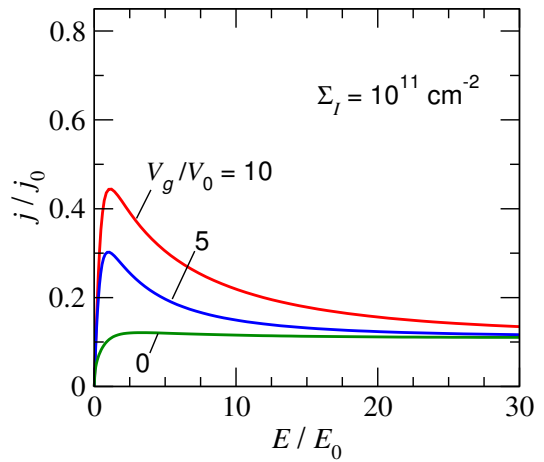}
\caption{The same as in Fig.~4,   but for $\Sigma_{I} = 10^{11}$~cm$^{-2}$.}
\label{F5}
\end{figure}

\begin{figure}[b]
\centering
\includegraphics[width=7.0cm]{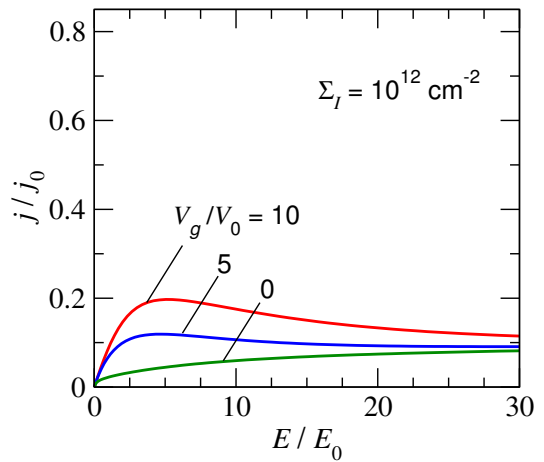}\\
\caption{The same as in Figs.~4 and 5, but for $\Sigma_{I} = 10^{12}$~cm$^{-2}$.}
\label{F6}
\end{figure}

Setting $\tau_0 \simeq 1$~ps  and $\tau_0^{inter} \simeq 1$~ps for room temperature, Eqs.~(17) and(18) yield the following estimates: $E_0 \simeq 86$~V/cm and $j_0 \simeq 3.49$~A/cm.
At $\kappa = 4$ and $W_g = 10^{-5} - 10^{-6}$~cm, the gate voltage is normalized by $V_0 \simeq 0.494 - 4.94 $~V.
Assuming that  $\Sigma_I = 10^{10}$~cm$^{-2}$  and using the peak values of the current density from Fig.~4, we obtain the following estimate for the FET transconductance $g = \Delta j/\Delta V_g$ : $g \simeq 42 - 420$~mS/mm. 
In the FETs with higher background dielectric constant   $\kappa $ and thinner gate layer $W_g$, the transconductance can be markedly larger.

In Fig.~7 we compare the current-voltage characteristics of two G-P channels both with  $\Sigma_I = 10^{11}$~cm$^{-2}$ but with different energy gaps ($\Delta = 900$~meV and $\Delta = 600$~meV).
As seen,  the current-voltage characteristics maxima increase with increasing  $\Delta$
and somewhat (weakly) shifts toward higher electric fields. In the P-layers with the number of the atomic layers
($N > 3$),  $\Delta^e \neq \Delta^h$ (actually $\Delta^e > \Delta^h$ ). The results obtained above  are qualitatively valid in such cases as well, but in the above formulas one needs to replace $\Delta/2 $ by $\Delta^h$. 

\begin{figure}[t]
\centering
\includegraphics[width=7.0cm]{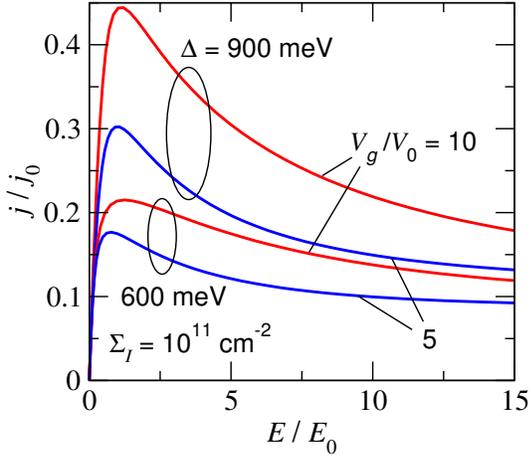}\\
\caption{Current-voltage characteristics for  $\Delta = 900$~meV (as in Fig.~5)
and $\Delta = 600$~meV at different $V_g/V_0$  ($\Sigma_I = 10^{11}$~cm$^{-3}$).}
\label{F7}
\end{figure}

\section{Discussion}

\subsection{Screening in the G-P channel}
The screening length in a semiconductor with  DoS$(\varepsilon)$, shown in Fig.~1(c),  is given by

\begin{equation}\label{Eq21}
l_{s}^{-1}=-\frac{2\pi {{e}^{2}}}{\kappa }\int\limits_{0}^{+\infty }{\rm DoS}( \varepsilon )\frac{d(f^e + f^h)}{d\varepsilon}d\varepsilon.
\end{equation}
 As the G- and P- layers are located close to each other (at the distance below the Fermi wavelength), their inverse screening lengths are additive. A simple evaluation with linear DoS in graphene and constant DoS in phosphorene leads us to
\begin{eqnarray}\label{Eq22} 
l_{s}^{-1}=\frac{e^2}{\kappa\hbar ^2}\biggl\{ \frac{4 T}{v_W^2}\displaystyle\ln\biggl[ \biggl( 1+e^{\mu^e/T} \biggr)\biggl( 1+e^{\mu^h/T} \biggr)\biggr] \nonumber\\
+2 N\sqrt{mM}
\biggl[\frac{1}{\displaystyle\exp\biggl(\frac{\Delta/2 - \mu^e}{T}\biggr) + 1}\nonumber\\
+ 
\frac{1}{\displaystyle\exp\biggl(-\frac{\Delta/2 + \mu^h}{T}\biggr) +1} \biggr]
  \biggr\}.
\end{eqnarray}
The first term comes from screening in G-layer, and the second one  from the P-layer. All the first factors in square brackets can be considered as effective relativistic mass in G-layers. 
At $\mu^e = \mu^h =0$, Eq.~(22) yields

\begin{equation}\label{eq23} 
l_s^{-1} = \frac{e^2}{\kappa\hbar^2}\biggl[8\ln 2 \frac{T}{v_W^2}  + 2N\sqrt{mM}\biggr]  > 8\alpha_gk_T\ln2,
\end{equation}
where $\alpha_g=(e^2/\hbar v_W\kappa)$ is the coupling constant for G-layers and $k_T = T/\hbar\,v_W$ is the characteristic  carrier wave number.  From Eq.~(23) we have the following estimate: 
$k_Tl_s \lesssim (8\alpha_g\ln2) ^{-1}$,
 At $\kappa = 4$,
one obtains  $k_Tl_s  \simeq  137/600\ln 2 \simeq 0.158\ll  1$. The dependence of the screening length $l_s $
on the quasi-Fermi energy $\mu = \mu^e$ (assuming that $\mu^h  =\hbar\omega(1-T/T_0)-\mu^e$)  calculated using Eq.~(23) is plotted in Fig.~8 ($\kappa = 4$ and $N = 3$). One can see that  an increase in $|\mu|$ leads to a decrease in the screening length $l_s$. 
The temperature dependence for $\mu = 0$ shown in the inset in Fig. 8 indicates that   $l_s$  decreases with $T$
except a narrow region near $T_0$, i.e., at low electric fields.  Relatively ineffective screening  at such fields somewhat affect the low-field conductivity [due to a distinction between the momentum dependences for the Coulomb scattering and the scattering on the neutral disorder (see Eq.~(3)], but is not important at high fields at which the NDC appears
because of the screening reinforcement.
 One can see that an increase in $T$ (the carrier heating) and $\mu$
  leads to a decrease in $l_s$ in comparison to the above estimate, so that  both inverse characteristic wavenumbers  $k_T^{-1} = \hbar\,v_W/T$
 and $k_{\mu}^{-1}  = \hbar\,v_W/\mu$ are markedly larger than $l_s$ (see the dashed lines in Fig.~8). This implies that the assumption of the  complete screening approximation is well-justified, particularly taking into account the contribution of the heavy carriers in the P-layer.

\begin{figure}[t]
\centering
	\includegraphics[width=7.0cm]{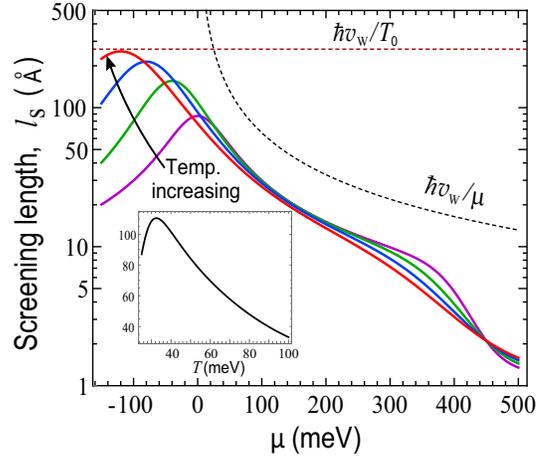}
\caption{Dependence of screening length in the G-P channel as a function of Fermi energy at various carrier effective temperatures increasing from $T = 25$~meV with 10~meV step. Number of P-layers $N=3$, background dielectric constant $\kappa=4$. Electrons and holes have different quasi-Fermi energies bound by $\mu^h = \hbar\omega(1-T/T_0)-\mu^e$. Inset: detailed temperature dependence of screening length at charge neutrality point $\mu=0$. Dashed lines correspond to $k_{T} ^{-1} |_{T=T_0}= \hbar\,v_W/T_0$
and $k_{\mu}^{-1} = \hbar\,v_W/\mu$  versus $\mu$.}
\end{figure}

\subsection{Mutual scattering of electrons and holes}

At the gate voltages corresponding to the states close to the  Dirac point, the mutual scattering of the electrons and holes in the G-layer  can affect the conductivity of the latter. However, due to special features of  the scattering of the carriers with the linear dispersion law~\cite{46,47}, such a scattering is similar to the scattering on uncharged and screened charged impurities, as well as acoustic phonons and defects. Hence, the  inclusion of  the inter-carrier scattering into the model should not markedly change the above results.

\subsection{Optical phonon heating}

In the case of relatively slow optical phonons decay processes, the heating of the optical phonon system
can be substantial, so that its effective temperature $T_0^{opt}$ can markedly exceed the thermal bath  temperature $T_0$ being close to the lattice temperature in the G-P channel (the temperature of acoustic phonons in this channel) and the carrier effective temperatures $T$ in the G-P channel.
 If such a decay is a "bottleneck", one can put $T_0^{opt} \simeq T$. In this case,  both $T$ and $T_0^{opt}$ 
 are determined by the lattice processes of the heat removal (characterized by their specific parameters and the device configuration).  In this case, in the Eqs.~(3) -(7) one needs to
 replace $T_0$ by $T_0^{opt} \simeq T$, so that Eq.~(11) yields $\mu^e + \mu^h \simeq 0$. One needs also to replace the left-hand side of  Eq.~(15) by the normalized value, $Q$, of the  heat flow from the G-P channel to the substrate and the gate layer.  For a simplified analysis  one can set $Q = K(T - T_0)(T/T_0)^q$, where $K$ characterizes the heat conductivity of the interfaces between the G-P-channel 
 and the surrounding layers the side contacts, as well as an efficiency of the heat removal via the side contacts and $q$ is a number.
 Assume fot the definiteness that $V_g = 0$ (i.e., $\mu^e = \mu^h = 0$ and set $q = 1$.  we arrive at the following dependence of the channel current  density on the common temperature of the carriers and the optical phonons
 in the G-P channel:
  
  \begin{equation}\label{eq24}
\frac{j}{j_0} \simeq \sqrt{\frac{K(T - T_0)}{1 + \displaystyle\frac{2\Sigma_N}{\Sigma_I}\frac{T}{T_0}
\displaystyle\exp\biggl(-\frac{\Delta}{2T}\biggr)}}.
\end{equation}
One can find that the latter $j - T$ relation exhibits a maximum at a certain temperature $T = T_{max} \sim
\Delta/[2\ln\Gamma(1 - \ln\Gamma/b)]$, where 
 $\Gamma =  (\Sigma_N\Delta/\Sigma_IT_0)$ and $b =\Delta/2T_0$.  At the parameters used in the above calculations, the latter estimate gives $T_{max} \simeq 48 - 82$~meV.  Due to a monotonic
increase in $T$ with increasing $E$, this implies that the pertinent current-voltage characteristic exhibits  the NDC
at sufficiently large $E$ when $T > T_{min}$ (at least when $q = 1$). 
 The more detailed consideration of this case requires a more accurate  model, that is
out of the scope of the present paper.

\subsection{Relaxation on substrate optical phonons}

If the electrons and holes effectively interact with the optical phonons of several types, say, with the G-layer optical phonons and the substrate optical phonons,  Eqs.~(9) - (11) should be properly generalized.
In particular, considering both interband and intraband transition, instead of Eq.~(11) governing the interband carrier balance one can arrive at

\begin{eqnarray}\label{eq25}
\frac{1}{\tau_0^{inter}}\exp\biggl( \frac{\mu^e + \mu^h-\hbar\omega_0}{T}\biggr) \nonumber\\
+\frac{1}{\tau_1^{inter}}\exp\biggl(\frac{\mu^e + \mu^h -\hbar \omega_1}{T}\biggr)\nonumber\\
 = \frac{1}{\tau_0^{inter}}\exp\biggl(- \frac{\hbar\omega_0}{T_0}\biggr) +\frac{1}{\tau_1^{inter}}\exp\biggl(- \frac{\hbar\omega_1}{T_0}\biggr) .
\end{eqnarray}
Here $\omega_1$ and $\tau_1^{inter}$ are the pertinent parameters for the substrate optical phonons.
As follows from Eq.~(25),  Eq.~(12) for $\mu^e +\mu^h$ should be replaced by the following:

\begin{equation}\label{eq26}
\frac{\mu^e + \mu^h}{T} =\ln\biggl[\frac{\tau_1^{inter}\exp(-\hbar\omega_0/T_0) +    \tau_0^{inter}\exp(-\hbar\omega_1/T_0)  }{\tau_1^{inter}\exp(-\hbar\omega_0/T) +    \tau_0^{inter}\exp(-\hbar\omega_1/T) }\biggr].
\end{equation}
If $\tau_1^{inter} \gg \tau_0^{inter} $ and $\omega_1 \sim \omega_0$, Eq.~(22) yields

\begin{eqnarray}\label{eq27}
\mu^e + \mu^h \simeq \hbar\omega_0\biggl(1 - \frac{T}{T_0}\biggr)\nonumber\\
+ T \biggl(\frac{\tau_0^{inter}}{\tau_1^{inter} }\biggr)\biggl[\exp\biggl(\frac{\hbar\omega_0 - \hbar\omega_1}{T}\biggr)
- \exp\biggl(\frac{\hbar\omega_0 - \hbar\omega_1}{T_0}\biggr)\biggr]\nonumber\\
\simeq \hbar\omega_0\biggl(1 - \frac{T}{T_0}\biggr)\biggl[ 1 
+ \biggl(\frac{\tau_0^{inter}}{\tau_1^{inter} }\biggr)\frac{(\omega_0 - \omega_1)}{\omega_0}\biggr].
\end{eqnarray}
Equation~(27) shows that the contribution of the substrate optical phonons can be roughly accounted for 
by a re-normalization of the quantity $\hbar\omega_0$.

In the opposite case, $\tau_1^{inter} \ll \tau_0^{inter} $ and $\omega_1 \ll \omega_0$,
from Eq.~(26) we obtain

\begin{equation}\label{eq28}
\mu^e + \mu^h \simeq \hbar\omega_1\biggl(1 - \frac{T}{T_0}\biggr) \ll \hbar\omega_0\biggl(1 - \frac{T}{T_0}\biggr) ,
\end{equation}
i.e, $\mu^e + \mu^h$ can be close to zero in a wide range of the effective temperature $T$.
Hence,   marked modifications of the above results can occur only in the case of  relatively strong interaction with the low energy substrate optical phonons.

\subsection{Possible applications of the G-P devices}

A steep decrease in the G-P channel conductivity with increasing effective temperature (see Fig.~2)  can be used for detection of the incident radiation in a wide spectral  range from terahertz (THz) to near infrared. The intraband
and interband  transitions caused by the absorbing photons result in the carrier heating and their redistribution between the G- and P- layers and, hence, in a decrease in the G-P-channel conductivity. This heating effect (providing the  negative photoconductivity) can  substantially stronger influence on the conductivity than that associated with the photogeneration of the extra carriers. The RST can markedly affect the response of the electron-hole plasma generated in the G-P channel by  ultrashort optical pulses increasing the efficiency of the photoconducting antennas
comprising  the G-P structures or their arrays. 

NDC  can lead to the instability of the electron-hole plasma in the G-P channel. This instability can be used for the generation of high-frequency oscillations in the device and the output microwave or THz radiation
(like in the Gunn diodes). 
The possibility of  THz operation can be limited by the speed of the carrier exchange between the G- and P-layers. Due to an effective  coupling of these layers and strong overlap of the pertinent wave function, one might expect that the inverse times  of the G-P carrier exchange fall into the terahertz range. 

Depending on the contact properties, the RST device with NDC could be used as a switch between low and high voltage states or a tunable current limiter. In a regime when the NDC leads to the formation of propagating high field domains profiling the P-layer enables the applications for logical circuits (including the non-Boolean logic circuits) and functional generators. Inserting a number of the RST devices into a THz waveguide will enable an operation in a highly efficient hybrid mode of quenching the domain formation for a high power integrated THz~source.

\section*{Conclusions}
We proposed the FETs with the G-P channel and calculated their characteristics using the developed device model.
We demonstrated that the carrier heating, in particular,  by the source-to-drain electric field leads to a substantial RST 
of the carriers between the G- and P-layers. As a result, the  population of the heavy carriers in the P-layer  strongly rises that results in a pronounced scattering reinforcement  of the light carriers in the G-layer and, consequently,  the drop of the G-P channel conductivity could lead to NDC. 
The FETs under consideration can be used for detection and generation of  electro-magnetic radiation and exhibit
nontrivial characteristics useful for the logical circuits and functional generators .

The authors are grateful to Prof. V. Vyurkov for valuable comments.
This work was supported by Japan Society for Promotion of Science (Grants Nos. 16H06361 and 16K14243),
Russian Science Foundation (Grant No. 14-29-00277), and
Russian Foundation for Basic Research (Grants Nos. 16-37-60110  and 18-07-01379). 
It was also partially supported by the RIEC Nation-Wide Collaborative Research Project, Japan.
The work at RPI was supported by Office of Naval Research (Project Monitor
Dr. Paul Maki).









\end{document}